\documentclass[prl,twocolumn,amsmath]{revtex4}
\usepackage{amsmath,amssymb,amsfonts,mathrsfs}
\usepackage{dsfont,amsthm}
\usepackage{graphicx}
\usepackage{pifont,color,url,booktabs,multirow}
\usepackage[utf8]{inputenc}  
\usepackage[T1]{fontenc}

\allowdisplaybreaks[4]
\def\be{\begin{equation}}
\def\ee{\end{equation}}
\def\ba{\begin{array}}
\def\ea{\end{array}}

\newcommand{\ket}[1]{|#1\rangle}

\begin{document}

\title{
Wave-particle-mixedness redistribution in Schwarzschild spacetime
}

\author{Sumeng Wang$^{1}$ and Xiaofen Huang$^{1, \dag}$}

\affiliation{
$^{1}$ School of Mathematics and Statistics, 
Hainan Normal University, 
Haikou, 571158, China\\
$^{\dag}$ Correspondence to: huangxf1206@163.com
}

\begin{abstract}
The redistribution of the wave feature, particle feature, and mixedness is investigated for  two-qubit isotropic states in Schwarzschild spacetime under Hawking radiation and environmental decoherence. 
It is shown that Hawking radiation changes their relative weights among different horizon regions rather than simply suppressing them. 
The analysis is further extended to phase damping, phase flip, and bit flip channels. 
Phase damping monotonically suppresses the wave feature and enhances mixedness, phase flip produces a symmetric death-and-revival behavior of the wave feature, and bit flip mainly reshapes the particle feature and mixedness through diagonal population redistribution. 
However, although Hawking radiation and channel noise affect the distribution of wave feature, particle feature, and mixedness in the subsystems, the triality relation among them still holds and remains unaffected by thermal and environmental noises.

\end{abstract}

\keywords{
Wave-particle-mixedness triality; 
Schwarzschild spacetime; 
Hawking radiation; 
Dirac field; 
quantum decoherence; 
quantum complementarity
}

\maketitle

\section{I. Introduction}

Wave-particle duality is one of the most fundamental manifestations of quantum mechanics and lies at the heart of Bohr's complementarity principle. In contrast to classical objects, a quantum system can exhibit both wave features and particle features, as most clearly illustrated by the double-slit experiment. This dual character is not only of foundational significance, but also plays an essential role in quantum metrology, quantum communication, quantum cryptography, quantum computation, and quantum information processing. Establishing a quantitative and physically transparent description of wave features and particle features has therefore remained a central problem in quantum foundations and quantum information science \cite{Bohr1928,Feynman1965,Wootters1979}.

Considerable effort has been devoted to formulating quantitative complementarity relations. An early and influential result was given by Greenberger and Yasin, who derived
\begin{equation}
P_{\rm GY}^{2}+W_{\rm GY}^{2}=1,
\label{GY_relation}
\end{equation}
where \(P_{\rm GY}\) denotes path predictability and \(W_{\rm GY}\) denotes interference visibility \cite{Greenberger1988}. This equality indicates that complete particle information and complete wave behavior cannot be simultaneously obtained. Englert later extended this idea by introducing which-path detectors and established the duality inequality
\begin{equation}
\mathcal{D}^{2}+\mathcal{V}^{2}\leq 1,
\label{Englert_relation}
\end{equation}
where \(\mathcal{D}\) and \(\mathcal{V}\) denote path distinguishability and fringe visibility, respectively \cite{Englert1996}. These results laid the foundation for the modern quantitative theory of complementarity. Subsequent studies generalized wave-particle duality to multibeam and multipath interferometers and revealed its close connection with quantum coherence, path information, and entanglement \cite{Durr2001,Englert2008,Peng2005,JakobBergou2007,JakobBergou2010}.

With the development of quantum information theory, wave-particle duality has gradually been reformulated from a broader resource-theoretic perspective. In such formulations, the population distribution in a reference basis characterizes the particle feature, while the off-diagonal terms of the density matrix characterize the wave feature. Duality relations built on quantum coherence have been established to link path distinguishability and quantum coherence, and relevant methods further elucidate the function of such coherence within wave-particle complementarity\cite{Bera2015,Bagan2016,Bagan2020}. More recently, the wave-particle-mixedness triality relation was proposed, in which the particle feature \(\mathcal{P}\), wave feature \(\mathcal{W}\), and mixedness \(\mathcal{M}\) satisfy
\begin{equation}
\mathcal{P}+\mathcal{W}+\mathcal{M}=1 .
\label{eq:intro_triality}
\end{equation}
This relation extends conventional wave-particle duality from pure state descriptions to mixed quantum states and provides a unified balance among population localization, off-diagonal contribution, and intrinsic statistical uncertainty \cite{Fu2022}. Therefore it offers a useful framework for analyzing situations in which wave-particle behavior and mixedness are simultaneously affected.

Relativistic quantum information provides a natural setting in which to test whether such a triality structure survives beyond the usual nonrelativistic or inertial description. Among various relativistic effects, Hawking radiation is particularly important because the presence of a black hole horizon makes the vacuum perceived by an outside observer behave as an effective thermal environment \cite{Hawking1974,Hawking1975,Hawking1976,Damour1976}. As a consequence, a quantum state near a black hole horizon can be decomposed into physically accessible and physically inaccessible sectors, and the partial trace over inaccessible modes generally changes the distribution of quantum features. During the past two decades, extensive studies have examined the influence of Hawking radiation on entanglement, quantum discord, coherence, Bell nonlocality, quantum steering, entropic uncertainty relations, and quantum teleportation fidelity \cite{Alsing2006,MartinMartinez2010,Wu2022,Wang2018,Li2022,Wang2019,Wang2024,Wu2025b,Wu2025f,Li2024,Mi2025,Mi2025b,Liu2024}. In parallel, environmental decoherence, usually modeled by noisy quantum channels such as phase damping, phase flip, and bit flip channels, provides another important source of redistribution or degradation of quantum properties \cite{Nielsen2010,Streltsov2017,Napoli2016,Horodecki2009}.

Despite these developments, the behavior of the wave feature, particle feature, and mixedness in curved spacetime has not been fully clarified. 
Existing studies on black-hole backgrounds have mainly focused on conventional quantum resources, whereas much less attention has been paid to how these three quantities are individually modified after the field modes are separated by the event horizon. 
This issue is nontrivial because the reduced states observed in Schwarzschild spacetime are obtained after tracing over causally inaccessible degrees of freedom, and the relative weights of the wave feature, particle feature, and mixedness may be redistributed among different horizon sectors. 
It is therefore important to determine how Hawking radiation changes these quantities and how their redistribution depends on the chosen noninertial bipartition.

 To address this problem, the wave-particle-mixedness framework is applied to a two-qubit isotropic state in Schwarzschild spacetime. 
After the field modes are separated into the exterior and interior sectors of the event horizon, four reduced density matrices for the noninertial bipartitions are obtained. 
For each reduced state, the wave feature, particle feature, and mixedness are derived explicitly. 
This formulation provides a direct way to identify how Hawking radiation redistributes these quantities between the physically accessible and physically inaccessible horizon sectors.

To further assess the robustness of this structure, environmental noise is introduced via three representative quantum channels, namely the phase-damping, phase-flip, and bit-flip channels.Since these channels affect the diagonal and off-diagonal density matrix elements differently, they provide complementary routes for modifying the wave feature, particle feature, and mixedness.The inclusion of these channels enables one to examine whether the triality relation remains valid not only in the Schwarzschild black hole background but also for noninertial reduced states subject to environmental noise.

The main purpose of this work is to reveal how the wave feature, particle feature, and mixedness are redistributed by Hawking radiation and by environmental decoherence in noninertial reduced states. 
In contrast to the conventional inertial setting, the present scenario incorporates two additional physical ingredients: the separation of the Dirac field modes across the event horizon and the environmental decoherence induced by noisy quantum channels. 
This setting makes it possible to compare how black-hole-induced mode decomposition and different noise channels modify the relative weights of the wave feature, particle feature, and mixedness.

This paper is organized as follows. 
In Sec.~II, we introduce the necessary preliminaries, 
including the quantization of the Dirac field in Schwarzschild 
spacetime, the wave-particle-mixedness triality relation, 
the initial two-qubit isotropic state, 
and the noisy quantum channels used in this work. 
In Sec.~III, we apply the triality framework to the reduced states 
generated by Hawking radiation and analyze the redistribution of 
the three quantities among different horizon sectors. 
In Sec.~IV, we examine the effects of the phase damping, 
phase flip, and bit flip channels, 
and compare their different redistribution mechanisms. 
Section~V summarizes the main conclusions and discusses possible 
extensions. 
The explicit forms of the reduced density matrices 
and the detailed analytical expressions are given in the Appendices.

\section{II. Preliminaries}
\label{sec:preliminaries}

\subsection{A. Schwarzschild spacetime and Dirac field modes}
\label{sec:Schwarzschild_background}

A system of natural units is adopted, in which the gravitational constant \(G\), speed of light \(c\), reduced Planck constant \(\hbar\), and Boltzmann constant \(k_{B}\) are set to unity, namely \(G=c=\hbar=k_{B}=1\). In these units, the Schwarzschild metric is written as
\begin{align}
ds^{2}
&=
-\left(1-\frac{2M}{R}\right)dt^{2}
+
\left(1-\frac{2M}{R}\right)^{-1}dR^{2}
\nonumber\\
&\quad
+
R^{2}d\theta^{2}
+
R^{2}\sin^{2}\theta d\phi^{2},
\label{eq:schwarzschild_metric}
\end{align}
where \(M\) denotes the mass of the black hole, \(R\) is the radial coordinate, \(t\) is the time coordinate, and \(\theta\) and \(\phi\) are the polar and azimuthal angles, respectively. The event horizon is located at \(R=2M\), which separates the exterior physically accessible region, denoted by Region I, from the interior physically inaccessible region, denoted by Region II.

In Schwarzschild spacetime, the massless Dirac equation is given by
\begin{equation}
\left[
\gamma^{a}e_{a}^{\mu}
\left(
\partial_{\mu}
+
\Gamma_{\mu}
\right)
\right]\psi
=
0,
\label{eq:dirac}
\end{equation}
where \(\gamma^{a}\) are the Dirac matrices in flat spacetime, \(e_{a}^{\mu}\) are the vierbein fields, and \(\Gamma_{\mu}\) is the spin connection. Substituting the Schwarzschild vierbein into Eq.~(\ref{eq:dirac}), one obtains
\begin{equation}
\begin{aligned}
&\frac{\gamma_{2}}{R}
\left(
\frac{\partial}{\partial\theta}
+
\frac{\cot\theta}{2}
\right)\Phi
+
\frac{\gamma_{3}}{R\sin\theta}
\frac{\partial\Phi}{\partial\phi}
-
\frac{\gamma_{0}}{\sqrt{1-\frac{2M}{R}}}
\frac{\partial \Phi}{\partial t}
\\
&\quad
+
\gamma_{1}
\sqrt{1-\frac{2M}{R}}
\left[
\frac{\partial}{\partial R}
+
\frac{1}{R}
+
\frac{M}{2R(R-2M)}
\right]\Phi
=
0,
\end{aligned}
\label{eq:dirac_explicit}
\end{equation}
where \(\gamma_i\) \((i=0,1,2,3)\) denote the Dirac gamma matrices.

Solving the Dirac equation near the event horizon gives two sets of positive-frequency outgoing fermionic modes,
\begin{equation}
\psi_{k}^{I+}
=
\xi e^{-i\omega u},
\qquad
\psi_{k}^{II+}
=
\xi e^{i\omega u},
\label{eq:positive_frequency_modes}
\end{equation}
where \(\psi_{k}^{I+}\) and \(\psi_{k}^{II+}\) correspond to the positive-frequency solutions outside and inside the event horizon, respectively. Here \(\omega\) is the monochromatic frequency of the Dirac field, \(\xi\) is a four-component Dirac spinor, and \(u=t-R_{*}\) is the retarded time. The tortoise coordinate is defined as
\begin{equation}
R_{*}
=
R
+
2M\ln\left(
\frac{R}{2M}-1
\right).
\label{eq:tortoise_coordinate}
\end{equation}

To remove the coordinate singularity at the event horizon, the Kruskal--Szekeres coordinates are introduced as
\begin{equation}
U
=
-
e^{-\frac{R_{*}}{2M}}
\left(
1-\frac{R}{2M}
\right)^{1/2}
e^{\frac{t}{4M}},
\label{eq:kruskal_U}
\end{equation}
\begin{equation}
V
=
e^{\frac{R_{*}}{2M}}
\left(
1-\frac{R}{2M}
\right)^{1/2}
e^{-\frac{t}{4M}}.
\label{eq:kruskal_V}
\end{equation}
Using the Damour--Ruffini analytic continuation method, the Kruskal modes \(\Phi_{k,I}^{+}\) and \(\Phi_{k,II}^{+}\) can be expressed in terms of the Schwarzschild modes as
\begin{align}
\Phi_{k,I}^{+}
&=
e^{-2\pi M\omega}
\psi_{-k,II}^{-}
+
e^{2\pi M\omega}
\psi_{k,I}^{+},
\label{eq:kruskal_mode_I}
\\
\Phi_{k,II}^{+}
&=
e^{-2\pi M\omega}
\psi_{-k,I}^{-}
+
e^{2\pi M\omega}
\psi_{k,II}^{+}.
\label{eq:kruskal_mode_II}
\end{align}

In Kruskal coordinates, the Dirac field can be expanded as
\begin{equation}
\begin{aligned}
\psi
&=
\int dk\,
[2\cosh(4\pi M\omega)]^{-1/2}
\Bigl[
\hat{c}_{k}^{II}\psi_{k,II}^{+}
+
\hat{d}_{-k}^{\dagger II}\psi_{-k,II}^{-}
\\
&\quad
+
\hat{c}_{k}^{I}\psi_{k,I}^{+}
+
\hat{d}_{-k}^{\dagger I}\psi_{-k,I}^{-}
\Bigr],
\end{aligned}
\label{eq:field_expansion}
\end{equation}
where \(\hat{c}_{k}\) and \(\hat{d}_{-k}^{\dagger}\) are the annihilation and creation operators associated with the corresponding fermionic modes.

The Bogoliubov transformation between Kruskal and Schwarzschild modes gives the vacuum and excited states of a Dirac mode in the black-hole background as
\begin{align}
\ket{0}_{K}
&=
\frac{1}{\sqrt{e^{-\omega/T}+1}}
\ket{0}_{I}\ket{0}_{II}
+
\frac{1}{\sqrt{e^{\omega/T}+1}}
\ket{1}_{I}\ket{1}_{II},
\label{eq:kruskal_vacuum}
\\
\ket{1}_{K}
&=
\ket{1}_{I}\ket{0}_{II}.
\label{eq:kruskal_excited}
\end{align}
Here \(T=1/(8\pi M)\) is the Hawking temperature. The Hawking parameter \(r\in[0,\pi/4]\) is introduced through
\(\cos r=1/\sqrt{e^{-\omega/T}+1}\) and
\(\sin r=1/\sqrt{e^{\omega/T}+1}\). In the following calculations, \(r_a\) and \(r_b\) are used to characterize the Hawking effects associated with Alice's and Bob's modes, respectively.

The above Schwarzschild--Kruskal mode relation shows that a field mode observed near the event horizon is naturally associated with two horizon sectors: the physically accessible exterior region and the physically inaccessible interior region. This black-hole-induced mode decomposition provides the basis for constructing the reduced density matrices used in the following sections. After tracing over the unobserved horizon modes, one obtains the reduced states \(\rho_{A_I B_I}\), \(\rho_{A_I B_{II}}\), \(\rho_{A_{II}B_I}\), and \(\rho_{A_{II}B_{II}}\), whose explicit forms are given in Appendix~A.

\subsection{B. Wave-particle-mixedness triality}

Following the wave-particle-mixedness framework proposed by 
Fu and Luo~\cite{Fu2022}, 
the particle feature, wave feature, and mixedness of a quantum state 
can be quantified in a unified way. 
For an \(n\)-dimensional quantum state \(\rho\) 
represented in the computational basis 
\(\{|i\rangle\}_{i=1}^{n}\), 
the particle feature describes the population distribution 
in the reference basis. 
It is defined as
\begin{equation}
\mathcal{P}(\rho)
=
\sum_{i=1}^{n}
\rho_{ii}^{2},
\label{eq:P}
\end{equation}
where \(\rho_{ii}\) denotes the diagonal element of density matrix of \(\rho\). 
A larger value of \(\mathcal{P}(\rho)\) indicates 
stronger population localization and hence a more pronounced 
particle feature.

The wave feature is determined by the off-diagonal elements 
of the density matrix. 
It is given by
\begin{equation}
\mathcal{W}(\rho)
=
\sum_{i\neq j}
|\rho_{ij}|^{2}.
\label{eq:W}
\end{equation}
It quantifies the contribution 
associated with off-diagonal coherence in the chosen basis.

The mixedness is measured by the linear entropy,
\begin{equation}
\mathcal{M}(\rho)
=
1-\mathrm{tr}(\rho^{2}),
\label{eq:M}
\end{equation}
which vanishes for a pure state and increases as the state 
becomes more mixed. 
With these definitions, the particle feature, wave feature, 
and mixedness satisfy the wave-particle-mixedness triality relation
\begin{equation}
\mathcal{P}(\rho)
+
\mathcal{W}(\rho)
+
\mathcal{M}(\rho)
=
1 .
\label{eq:triality}
\end{equation}
This relation, originally introduced in Ref.~\cite{Fu2022}, 
gives a unified balance among population localization, 
off-diagonal contribution, and statistical mixedness.

\section{III. Wave-particle-mixedness redistribution under Hawking radiation}
\label{sec:pure_hawking_triality}

Based on the wave-particle-mixedness framework introduced in Sec.~II, the isotropic state in Schwarzschild spacetime is analyzed below.
The aim is to clarify how Hawking radiation redistributes the wave feature, particle feature, and mixedness across different horizon region.

The initial state is taken to be a two-qubit isotropic state shared by Alice and Bob,
\begin{equation}
\rho_{\mathrm{iso}}
=
\frac{1-\alpha}{4}
I\otimes I
+
\alpha
|\psi^{+}\rangle\langle\psi^{+}|,
\label{rhoiso}
\end{equation}
where 
\(|\psi^{+}\rangle
=
\frac{1}{\sqrt{2}}\bigl(|01\rangle+|10\rangle\bigr)\)
and \(\alpha\in[0,1]\). 
The parameter \(\alpha\) specifies the weight of the Bell-state component and therefore determines the initial distribution of the wave feature, particle feature, and mixedness.

Alice and Bob are taken to be freely falling observers approaching the Schwarzschild black hole.
After the Kruskal modes are separated into the physically accessible exterior region and the physically inaccessible interior region of the event horizon, the original bipartite state gives rise to four reduced density matrices:
$
\rho_{A_I B_I},$
$\rho_{A_I B_{II}},$
$\rho_{A_{II}B_I},$
$\rho_{A_{II}B_{II}} .
$
Their explicit matrix forms are given in Appendix~A. 
Using the definitions in Eqs.~(\ref{eq:P})--(\ref{eq:M}), 
the wave feature, particle feature, and mixedness are evaluated 
for each reduced state as shown in Eqs.(\ref{eq:W_reduced_all}), (\ref{eq:P_reduced_all}) and (\ref{eq:M_reduced_all}).
The results are then substituted into Eq.~(\ref{eq:triality}) 
to verify whether the triality balance is preserved under 
Hawking radiation.

\begin{widetext}

For the reduced states $
\rho_{A_I B_I},$
$\rho_{A_I B_{II}},$
$\rho_{A_{II}B_I},$
$\rho_{A_{II}B_{II}} 
$,  
the wave features are
\begin{equation}
\begin{alignedat}{2}
\mathcal{W}_{A_I B_I}
&=
\frac{\alpha^2}{2}
\cos^2 r_a
\cos^2 r_b ,
\qquad
&
\mathcal{W}_{A_I B_{II}}
&=
\frac{\alpha^2}{2}
\cos^2 r_a
\sin^2 r_b ,
\\[7pt]
\mathcal{W}_{A_{II}B_I}
&=
\frac{\alpha^2}{2}
\sin^2 r_a
\cos^2 r_b ,
\qquad
&
\mathcal{W}_{A_{II}B_{II}}
&=
\frac{\alpha^2}{2}
\sin^2 r_a
\sin^2 r_b .
\end{alignedat}
\label{eq:W_reduced_all}
\end{equation}

The  particle features are
\begin{equation}
\begin{aligned}
\mathcal{P}_{A_I B_I}
&=
\frac{1}{4}
\Bigl[
1
+
\sin^4 r_a
+
\sin^4 r_b
+
\bigl(
\sin^2 r_a
\sin^2 r_b
-
\alpha
\cos^2 r_a
\cos^2 r_b
\bigr)^2
\Bigr],
\\[8pt]
\mathcal{P}_{A_I B_{II}}
&=
\frac{1}{4}
\Bigl[
1
+
\sin^4 r_a
+
\cos^4 r_b
+
\bigl(
\alpha
\cos^2 r_a
\sin^2 r_b
-
\sin^2 r_a
\cos^2 r_b
\bigr)^2
\Bigr],
\\[8pt]
\mathcal{P}_{A_{II}B_I}
&=
\frac{1}{4}
\Bigl[
1
+
\cos^4 r_a
+
\sin^4 r_b
+
\bigl(
\alpha
\sin^2 r_a
\cos^2 r_b
-
\cos^2 r_a
\sin^2 r_b
\bigr)^2
\Bigr],
\\[8pt]
\mathcal{P}_{A_{II}B_{II}}
&=
\frac{1}{4}
\Bigl[
1
+
\cos^4 r_a
+
\cos^4 r_b
+
\bigl(
\cos^2 r_a
\cos^2 r_b
-
\alpha
\sin^2 r_a
\sin^2 r_b
\bigr)^2
\Bigr].
\end{aligned}
\label{eq:P_reduced_all}
\end{equation}

The mixednesses obtained from the linear entropy are
\begin{equation}
\begin{aligned}
\mathcal{M}_{A_I B_I}
&=
\frac{1}{4}
\Bigl[
3
-
\bigl(
\sin^2 r_a
\sin^2 r_b
-
\alpha
\cos^2 r_a
\cos^2 r_b
\bigr)^2
-
\sin^4 r_a
-
\sin^4 r_b
-
2\alpha^2
\cos^2 r_a
\cos^2 r_b
\Bigr],
\\[8pt]
\mathcal{M}_{A_I B_{II}}
&=
\frac{1}{4}
\Bigl[
3
-
\bigl(
\alpha
\cos^2 r_a
\sin^2 r_b
-
\sin^2 r_a
\cos^2 r_b
\bigr)^2
-
\sin^4 r_a
-
\cos^4 r_b
-
2\alpha^2
\cos^2 r_a
\sin^2 r_b
\Bigr],
\\[8pt]
\mathcal{M}_{A_{II}B_I}
&=
\frac{1}{4}
\Bigl[
3
-
\bigl(
\alpha
\sin^2 r_a
\cos^2 r_b
-
\cos^2 r_a
\sin^2 r_b
\bigr)^2
-
\cos^4 r_a
-
\sin^4 r_b
-
2\alpha^2
\sin^2 r_a
\cos^2 r_b
\Bigr],
\\[8pt]
\mathcal{M}_{A_{II}B_{II}}
&=
\frac{1}{4}
\Bigl[
3
-
\bigl(
\cos^2 r_a
\cos^2 r_b
-
\alpha
\sin^2 r_a
\sin^2 r_b
\bigr)^2
-
\cos^4 r_a
-
\cos^4 r_b
-
2\alpha^2
\sin^2 r_a
\sin^2 r_b
\Bigr].
\end{aligned}
\label{eq:M_reduced_all}
\end{equation}

\end{widetext}

The expressions above explicitly show how Hawking radiation redistributes the wave feature, particle feature, and mixedness among different horizon regions.
As a consistency check, substituting the corresponding expressions of 
\(\mathcal{W}\), 
\(\mathcal{P}\), 
and 
\(\mathcal{M}\) 
into Eq.~(\ref{eq:triality}) yields unity for 
\(\rho_{A_I B_I}\), 
\(\rho_{A_I B_{II}}\), 
\(\rho_{A_{II}B_I}\), 
and 
\(\rho_{A_{II}B_{II}}\), respectively.
Thus, the Hawking effect modifies the distribution of the wave feature, particle feature, and mixedness without violating their total balance.

Fig.~\ref{fig:hawking_WPM_group} presents the wave feature, 
particle feature, and mixedness as functions of the Hawking parameter 
\(r\) under the symmetric condition \(r_a=r_b=r\). 
As shown in panel (a), 
the wave feature of \(\rho_{A_I B_I}\) decreases monotonically 
with increasing \(r\), whereas that of \(\rho_{A_I B_{II}}\) 
increases over the plotted range. 
For \(\rho_{A_{II}B_{II}}\), the wave feature grows more rapidly 
as the contribution from the physically inaccessible region becomes stronger.

\begin{figure*}[!t]
\centering

\makebox[\textwidth][c]{%
\includegraphics[width=1.04\textwidth]{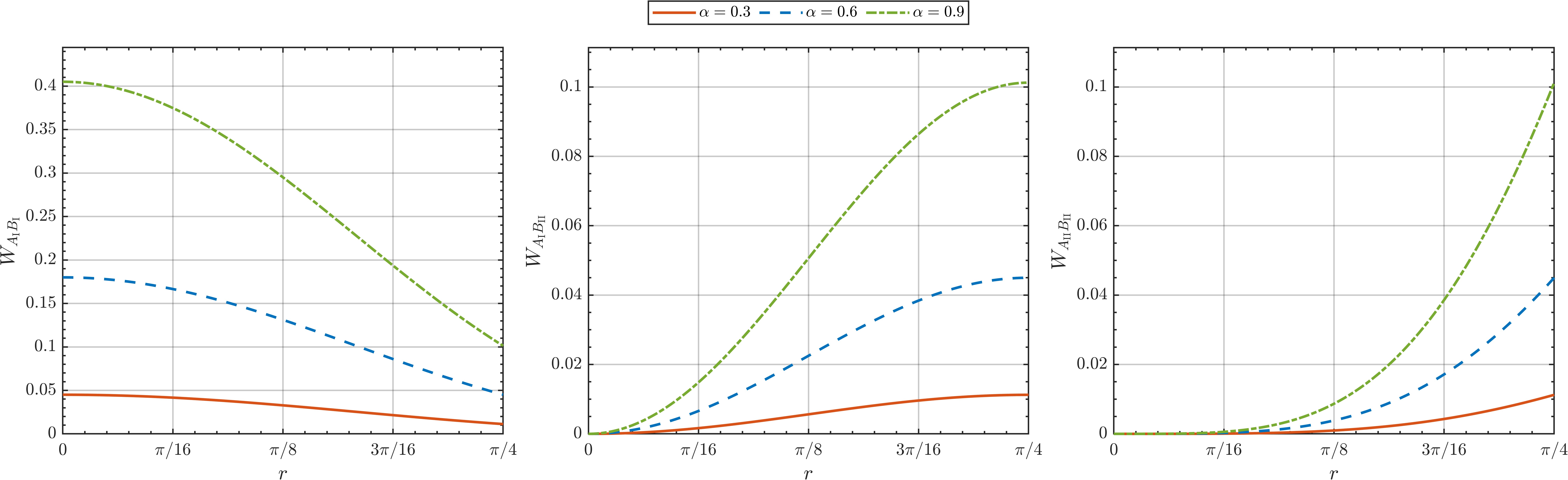}
}

\vspace{1mm}
\centerline{\textbf{(a)}}

\vspace{3mm}

\makebox[\textwidth][c]{%
\includegraphics[width=1.04\textwidth]{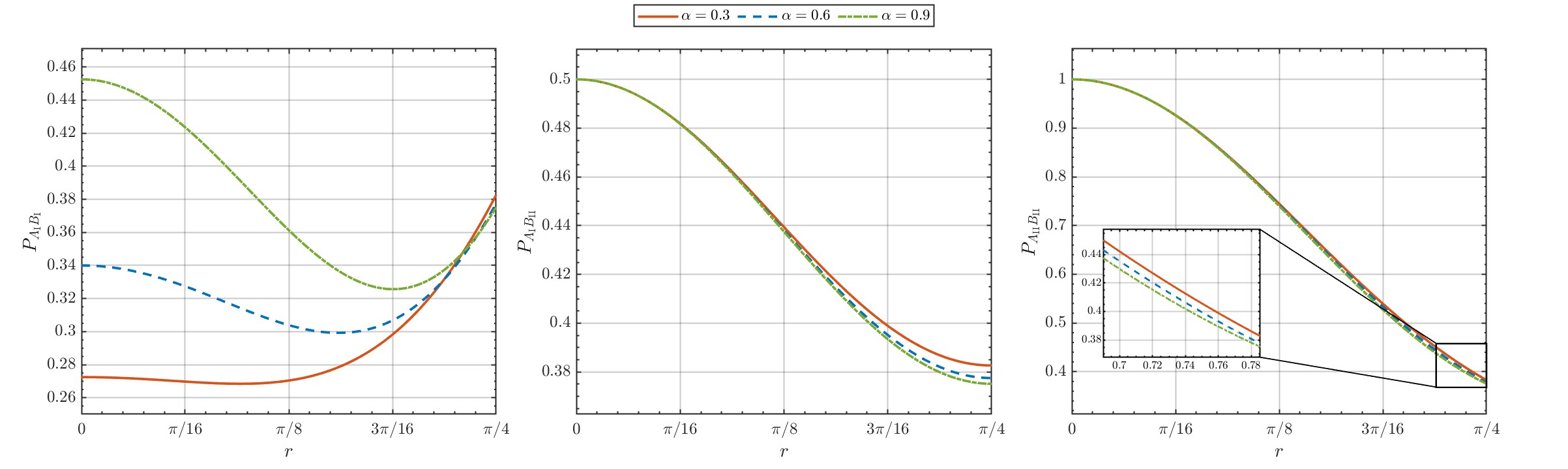}
}

\vspace{1mm}
\centerline{\textbf{(b)}}

\vspace{3mm}

\makebox[\textwidth][c]{%
\includegraphics[width=1.04\textwidth]{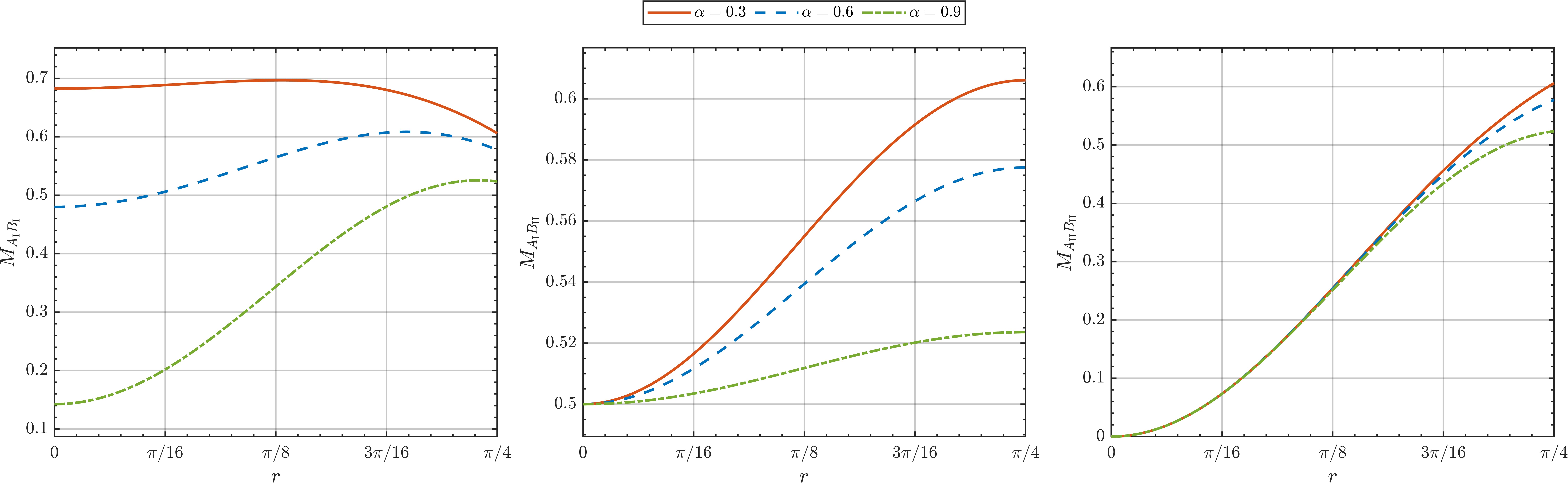}
}

\vspace{1mm}
\centerline{\textbf{(c)}}

\vspace{2mm}

\caption{
Wave feature \(\mathcal{W}\), particle feature \(\mathcal{P}\), and mixedness \(\mathcal{M}\) versus the Hawking parameter \(r\) for \(r_a=r_b=r\). Panels (a)--(c) show \(\mathcal{W}\), \(\mathcal{P}\), and \(\mathcal{M}\), respectively. The three subplots in each panel correspond to \(\rho_{A_I B_I}\), \(\rho_{A_I B_{II}}\), and \(\rho_{A_{II}B_{II}}\). The curves are for \(\alpha=0.3,0.6,0.9\). The state \(\rho_{A_{II}B_I}\) is omitted because it coincides with \(\rho_{A_I B_{II}}\) when \(r_a=r_b\).
}
\end{figure*}

\label{fig:hawking_WPM_group}

Fig.~\ref{fig:hawking_WPM_group}(b) presents the corresponding 
particle feature. 
For the physically accessible reduced state 
\(\rho_{A_I B_I}\), 
\(\mathcal{P}\) varies nonmonotonically with increasing \(r\). 
This behavior arises from the competition between the decreasing 
physically accessible contribution and the population redistribution 
induced by the Hawking-mode decomposition. 
For \(\rho_{A_I B_{II}}\), 
\(\mathcal{P}\) decreases as \(r\) increases, whereas for 
\(\rho_{A_{II}B_{II}}\), 
it is mainly governed by the population structure of the physically 
inaccessible region. 
The parameter \(\alpha\) controls the separation among the curves by 
changing the weight of the Bell state component.

Fig.~\ref{fig:hawking_WPM_group}(c) presents the mixedness. 
In general, \(\mathcal{M}\) exhibits a trend complementary to those 
of the wave feature and particle feature. 
For \(\rho_{A_I B_I}\), 
the suppression of the wave feature is accompanied by an increase 
in mixedness, indicating that the reduced state becomes more 
statistically uncertain as the off-diagonal contribution decreases. 
For the reduced states involving Region-II modes, 
\(\mathcal{M}\) increases as the contribution from the physically 
inaccessible sector becomes stronger. 
In the high temperature limit \(r_a=r_b=\pi/4\), 
one has \(\cos^2 r=\sin^2 r=1/2\), 
so the exterior and interior contributions become comparable. 
Thus, Hawking radiation does not simply suppress the wave feature, 
particle feature, and mixedness; instead, it redistributes them 
among different horizon sectors.

\section{IV. Wave-particle-mixedness dynamics under environmental noise}
\label{sec:decoherence_triality}

To examine how environmental noise further modifies the wave feature, 
particle feature, and mixedness in noninertial reduced states, 
three representative noise channels acting on a single qubit are considered: 
the phase damping (PD), phase flip (PF), and bit flip (BF) channels. 
Their Kraus operators are listed in Table~\ref{tab:noise_channels}, 
where \(k\in[0,1]\) denotes the noise strength.

\begin{table}[h]
\centering
\small
\renewcommand{\arraystretch}{1.35}
\resizebox{\linewidth}{!}{
\begin{tabular}{c c}
\hline
Channel & Kraus operators \\
\hline

PD &
$
K_0^{\mathrm{PD}}
=
\begin{pmatrix}
1 & 0 \\
0 & \sqrt{1-k}
\end{pmatrix},
\quad
K_1^{\mathrm{PD}}
=
\begin{pmatrix}
0 & 0 \\
0 & \sqrt{k}
\end{pmatrix}
$
\\[10pt]

PF &
$
K_0^{\mathrm{PF}}
=
\sqrt{1-k}\,I,
\quad
K_1^{\mathrm{PF}}
=
\sqrt{k}\,\sigma_z
$
\\[8pt]

BF &
$
K_0^{\mathrm{BF}}
=
\sqrt{1-k}\,I,
\quad
K_1^{\mathrm{BF}}
=
\sqrt{k}\,\sigma_x
$
\\
\hline
\end{tabular}
}
\caption{
Kraus representations of the phase damping, 
phase flip, and bit flip channels. 
The parameter \(k\) is the damping coefficient.
}
\label{tab:noise_channels}
\end{table}

Let
$
X
\in
\{
A_I B_I,
A_I B_{II},
A_{II}B_I,
A_{II}B_{II}
\}
$
denote one of the four two-body reduced states obtained after 
the Schwarzschild mode decomposition. 
For a given channel 
\(\mu\in\{\mathrm{PD},\mathrm{PF},\mathrm{BF}\}\), 
the noisy reduced state is written as
\begin{equation}
\rho_X^{\mu}
=
\mathcal{E}_{\mu}(\rho_X)
=
\sum_{\ell}
K_{\ell}^{\mu}
\rho_X
K_{\ell}^{\mu\dagger},
\qquad
\sum_{\ell}
K_{\ell}^{\mu\dagger}
K_{\ell}^{\mu}
=
I .
\label{eq:channel_action_general}
\end{equation}
Here \(K_{\ell}^{\mu}\) denotes the Kraus operator of 
the corresponding noisy channel. 
The explicit matrix forms of the noisy reduced states 
are given in Appendix~B.

This section extends the previous analysis to noisy noninertial 
reduced states. 
After the Hawking-induced mode decomposition, the reduced states 
\(\rho_{A_I B_I}\), 
\(\rho_{A_I B_{II}}\), 
\(\rho_{A_{II}B_I}\), 
and 
\(\rho_{A_{II}B_{II}}\) 
are further subjected to environmental noise. 
The aim is to determine how the wave feature, particle feature, 
and mixedness are modified by the phase damping, phase flip, 
and bit flip channels introduced above. 
The triality relation is used only as a consistency check for the 
quantities calculated from the noisy reduced density matrices.

For each noisy channel 
\(\mu\in\{\mathrm{PD},\mathrm{PF},\mathrm{BF}\}\), 
the reduced state is evolved according to 
Eq.~(\ref{eq:channel_action_general}). 
The resulting noisy density matrices are given in Appendix~B. 
The corresponding particle feature, wave feature, and mixedness 
are calculated from Eqs.~(\ref{eq:P})--(\ref{eq:M}), 
with the mixedness obtained independently from the linear entropy.

\setlength{\dbltextfloatsep}{10pt}
\setlength{\dblfloatsep}{8pt}
\setlength{\textfloatsep}{10pt}
\setlength{\floatsep}{8pt}
\setlength{\intextsep}{8pt}

\subsection{A. Phase damping channel}

The phase damping channel suppresses the off-diagonal elements 
of the reduced density matrix, 
while leaving the diagonal populations unchanged. 
For the reduced states considered here, 
this channel introduces the factor \((1-k)^2\) 
into the wave feature, 
whereas the particle feature has no explicit dependence on 
the damping coefficient \(k\). 
The mixedness is evaluated independently from the linear entropy.

The noisy reduced states generated by the PD channel, 
together with their corresponding wave feature, 
particle feature, and mixedness, 
are listed in Appendix~B. 
Eqs.~(\ref{eq:PD_W_all})--(\ref{eq:PD_M_all}) 
show that the phase damping channel modifies the wave feature and 
mixedness through the factor \((1-k)^2\), whereas the particle 
feature remains independent of the noise strength \(k\). 

Under the symmetric condition \(r_a=r_b=r\), 
the reduced states 
\(\rho^{\mathrm{PD}}_{A_I B_{II}}\) 
and 
\(\rho^{\mathrm{PD}}_{A_{II}B_I}\) 
become identical. 
Therefore, only 
\(\rho^{\mathrm{PD}}_{A_I B_I}\), 
\(\rho^{\mathrm{PD}}_{A_I B_{II}}\), 
and 
\(\rho^{\mathrm{PD}}_{A_{II}B_{II}}\) 
are presented in the following figures. Fig.~\ref{fig:PD_WM_all} displays the wave feature and mixedness 
under the phase damping channel. 
Panel (a) corresponds to the wave feature, while panel (b) corresponds 
to the mixedness, both plotted as functions of the damping coefficient \(k\). 
Since the particle feature has no explicit \(k\)-dependence under the 
phase damping channel, it is not plotted here. 

\begin{figure*}[!t]
\centering

\makebox[\textwidth][c]{%
\includegraphics[width=1.04\textwidth]{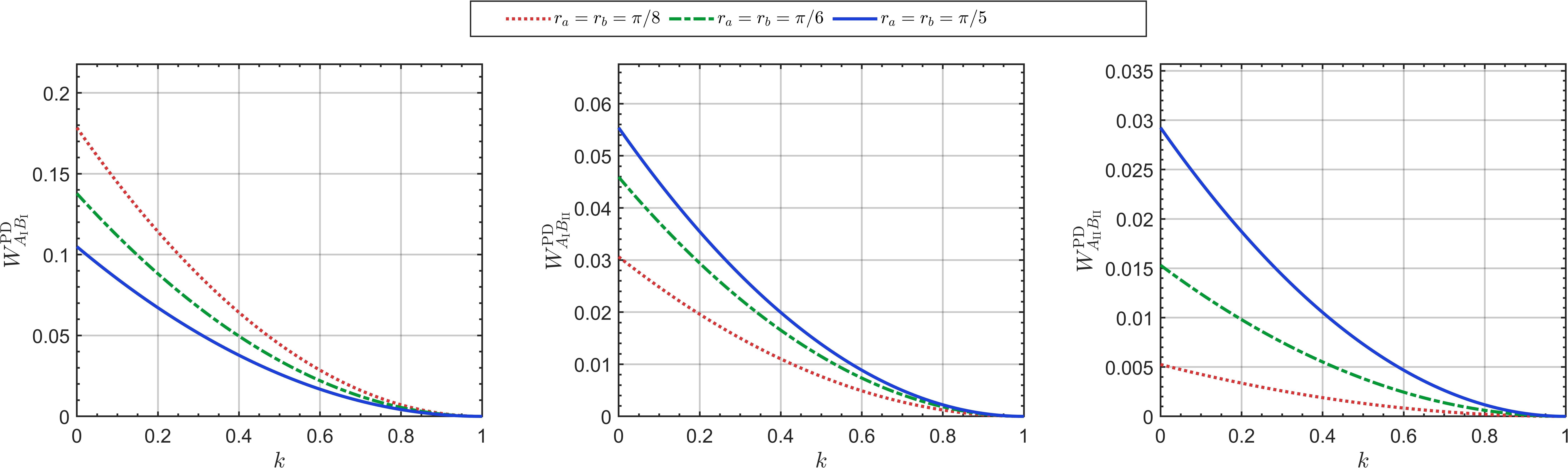}
}

\vspace{1mm}
\centerline{\textbf{(a)}}

\vspace{3mm}

\makebox[\textwidth][c]{%
\includegraphics[width=1.04\textwidth]{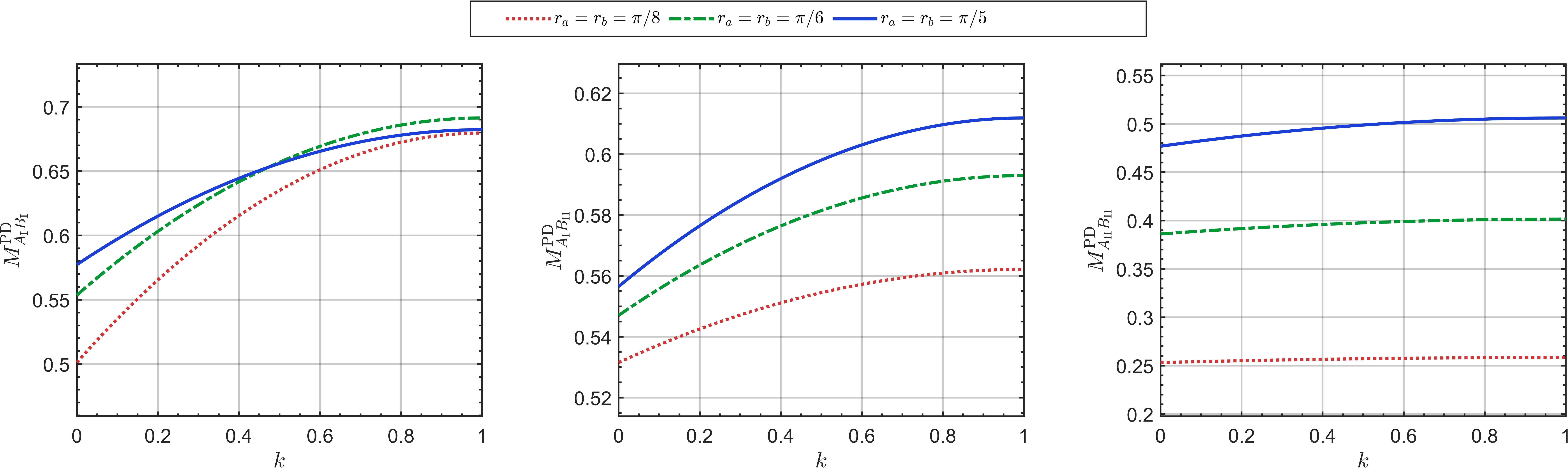}
}

\vspace{1mm}
\centerline{\textbf{(b)}}

\vspace{2mm}

\caption{
Wave feature and mixedness under the phase damping channel 
for \(\alpha=0.7\) and \(r_a=r_b=r\). 
Panel (a) shows the wave feature 
\(\mathcal{W}^{\mathrm{PD}}\), 
while panel (b) shows the mixedness 
\(\mathcal{M}^{\mathrm{PD}}\), 
both as functions of the damping coefficient \(k\). 
In each panel, the three subplots correspond to 
\(\rho^{\mathrm{PD}}_{A_I B_I}\), 
\(\rho^{\mathrm{PD}}_{A_I B_{II}}\), 
and 
\(\rho^{\mathrm{PD}}_{A_{II}B_{II}}\). 
The three curves represent 
\(r=\pi/8\), 
\(r=\pi/6\), 
and 
\(r=\pi/5\). 
The particle feature is not plotted because it has no explicit 
\(k\)-dependence. 
For \(r_a=r_b=r\), 
\(\rho^{\mathrm{PD}}_{A_{II}B_I}\) 
has the same behavior as 
\(\rho^{\mathrm{PD}}_{A_I B_{II}}\).
}
\label{fig:PD_WM_all}
\end{figure*}

As shown by subplot (a) in Fig.~(\ref{fig:PD_WM_all}), the wave feature of 
\(\rho^{\mathrm{PD}}_{A_I B_I}\), 
\(\rho^{\mathrm{PD}}_{A_I B_{II}}\), 
and 
\(\rho^{\mathrm{PD}}_{A_{II}B_{II}}\) 
decreases monotonically as \(k\) increases. 
At \(k=1\), the factor \((1-k)^2\) vanishes, so the wave feature also 
vanishes, corresponding to the sudden death of the wave feature under 
complete phase damping. 
The Hawking parameter \(r\) mainly determines the initial amplitude of 
the curves, whereas the damping coefficient \(k\) directly controls the 
suppression of the wave feature.

Fig.~\ref{fig:PD_WM_all}(b) presents the corresponding mixedness. 
In contrast to the wave feature, the mixedness increases with the 
damping coefficient \(k\). 
This behavior follows from Eq.~(\ref{eq:PD_M_all}), where the 
\(k\)-dependent the wave feature contribution enters with a negative sign. 
Therefore, the suppression of the wave feature is accompanied by an 
increase in mixedness. 
The variation is more pronounced when the initial wave feature amplitude 
is larger, whereas it becomes weaker for reduced states with smaller 
off-diagonal contributions.

In summary, the phase damping channel produces a monotonic redistribution 
between the wave feature and mixedness, while the particle feature remains 
independent of the damping coefficient. 
At \(k=1\), the wave feature vanishes, whereas the mixedness reaches its 
maximum value at the same point. 
After the wave feature, particle feature, and mixedness are obtained 
independently from the noisy reduced states, the calculated quantities also 
satisfy Eq.~(\ref{eq:triality}) after substitution.

\subsection{B. Phase flip channel}

The phase flip channel leaves the diagonal elements of density matrix  unchanged, 
whereas each nonzero off-diagonal matrix element is multiplied by the 
factor \((1-2k)\). 
Consequently, the wave feature contains the factor \((1-2k)^2\), 
while the particle feature has no explicit dependence on the damping 
coefficient \(k\). 
The mixedness is evaluated independently from the linear entropy of the 
corresponding noisy reduced density matrix. 
The noisy reduced states generated by the PF channel, together with the 
corresponding wave feature, particle feature, and mixedness, are given 
in Appendix~B.

\begin{figure*}[!t]
\centering

\makebox[\textwidth][c]{%
\includegraphics[width=1.04\textwidth]{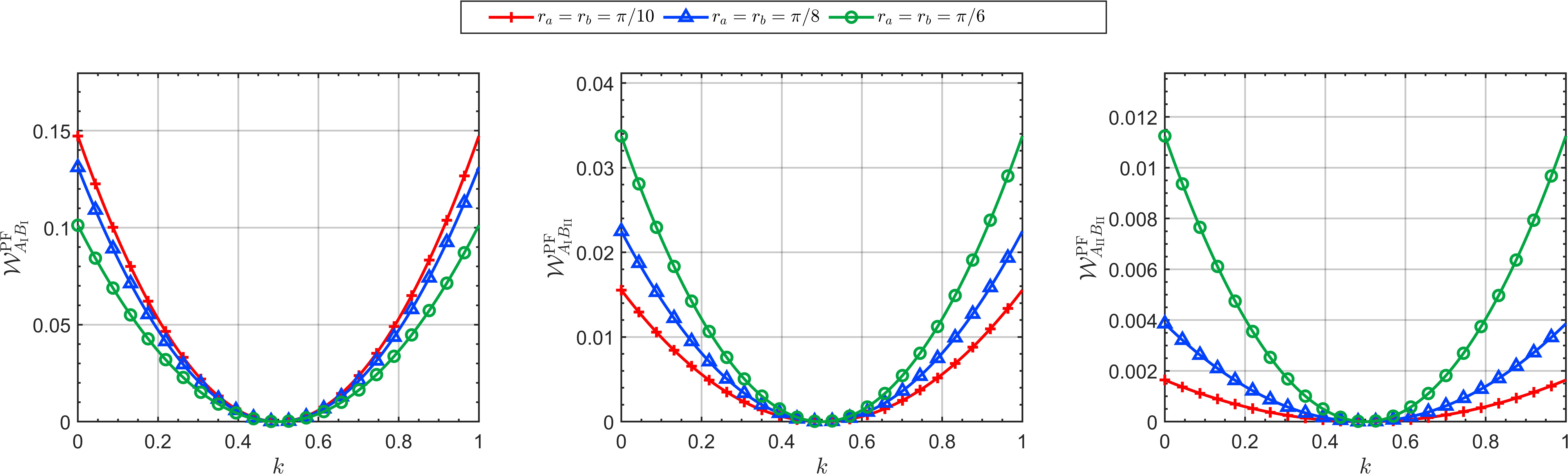}
}

\vspace{1mm}
\centerline{\textbf{(a)}}

\vspace{3mm}

\makebox[\textwidth][c]{%
\includegraphics[width=1.04\textwidth]{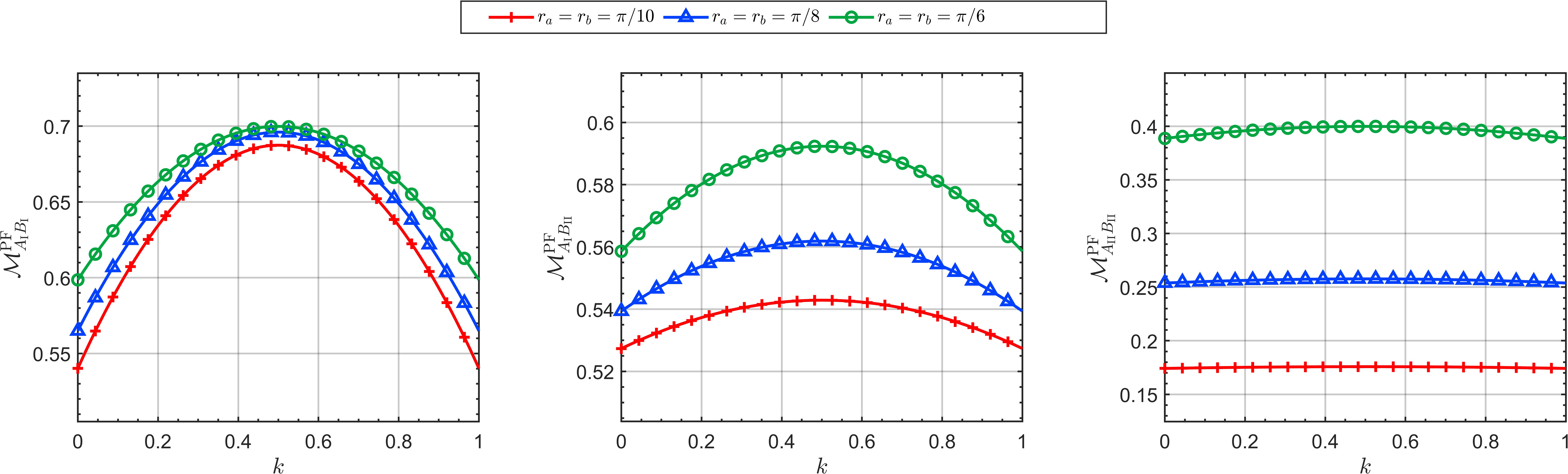}
}

\vspace{1mm}
\centerline{\textbf{(b)}}

\vspace{2mm}

\caption{
Wave feature and mixedness under the phase flip channel 
for \(\alpha=0.6\) and \(r_a=r_b=r\). 
Panel (a) shows the wave feature 
\(\mathcal{W}^{\mathrm{PF}}\), 
while panel (b) shows the mixedness 
\(\mathcal{M}^{\mathrm{PF}}\), 
both as functions of the damping coefficient \(k\). 
In each panel, the three subplots correspond to 
\(\rho^{\mathrm{PF}}_{A_I B_I}\), 
\(\rho^{\mathrm{PF}}_{A_I B_{II}}\), 
and 
\(\rho^{\mathrm{PF}}_{A_{II}B_{II}}\). 
The three curves represent 
\(r=\pi/10\), 
\(r=\pi/8\), 
and 
\(r=\pi/6\). 
The particle feature is not plotted because it has no explicit 
\(k\)-dependence. 
For \(r_a=r_b=r\), 
\(\rho^{\mathrm{PF}}_{A_{II}B_I}\) 
has the same behavior as 
\(\rho^{\mathrm{PF}}_{A_I B_{II}}\).
}
\label{fig:PF_WM_all}
\end{figure*}

Eqs.~(\ref{eq:PF_W_all_C})--(\ref{eq:PF_M_all_C}) show that 
the PF channel modifies the wave feature and mixedness through the 
factor \((1-2k)^2\), whereas the particle feature remains independent 
of the damping coefficient \(k\). 
Therefore, only the wave feature and mixedness are displayed in 
Fig.~\ref{fig:PF_WM_all}. 
Under the symmetric condition \(r_a=r_b=r\), the reduced states 
\(\rho^{\mathrm{PF}}_{A_I B_{II}}\) 
and 
\(\rho^{\mathrm{PF}}_{A_{II}B_I}\) 
exhibit identical behavior, so only one of them is shown.

Fig.~\ref{fig:PF_WM_all}(a) presents the wave feature under the 
phase flip channel as a function of the damping coefficient \(k\). 
For 
\(\rho^{\mathrm{PF}}_{A_I B_I}\), 
\(\rho^{\mathrm{PF}}_{A_I B_{II}}\), 
and 
\(\rho^{\mathrm{PF}}_{A_{II}B_{II}}\), 
the wave feature is symmetric about \(k=1/2\). 
As \(k\) increases from \(0\) to \(1/2\), the wave feature is gradually 
suppressed to zero; it then increases again as \(k\) further increases 
from \(1/2\) to \(1\). 
This death-and-revival behavior originates from the factor 
\((1-2k)^2\) in Eq.~(\ref{eq:PF_W_all_C}). 
The Hawking parameter \(r\) mainly changes the amplitudes of the curves. 
For \(\rho^{\mathrm{PF}}_{A_I B_I}\), the amplitude decreases with 
increasing \(r\). 
By contrast, for 
\(\rho^{\mathrm{PF}}_{A_I B_{II}}\) and 
\(\rho^{\mathrm{PF}}_{A_{II}B_{II}}\), 
the amplitude is enhanced as the Region-II contribution becomes larger.

Fig.~\ref{fig:PF_WM_all}(b) presents the corresponding mixedness. 
The mixedness also exhibits a symmetric dependence on the damping 
coefficient \(k\). 
In contrast to the wave feature, it increases as \(k\) varies from 
\(0\) to \(1/2\), reaches its maximum when the wave feature vanishes, 
and then decreases symmetrically for \(k>1/2\). 
This opposite behavior follows from Eq.~(\ref{eq:PF_M_all_C}), where 
the \(k\)-dependent wave feature term enters the mixedness with a 
negative sign. 
The variation is more pronounced for 
\(\rho^{\mathrm{PF}}_{A_I B_I}\) 
and 
\(\rho^{\mathrm{PF}}_{A_I B_{II}}\), 
but becomes less visible for 
\(\rho^{\mathrm{PF}}_{A_{II}B_{II}}\). 
Since Eq.~(\ref{eq:PF_P_all_C}) shows that the particle feature is 
independent of \(k\), it is not shown in the figure. 
Therefore, the PF channel mainly redistributes the wave-particle-mixedness 
distribution between the wave feature and mixedness.

Substituting 
Eqs.~(\ref{eq:PF_W_all_C}), 
(\ref{eq:PF_P_all_C}), 
and 
(\ref{eq:PF_M_all_C}) 
into Eq.~(\ref{eq:triality}) also gives 
\(\mathcal{P}(\rho)+\mathcal{W}(\rho)+\mathcal{M}(\rho)=1\). 
Therefore, the PF channel induces a symmetric redistribution between 
the wave feature and mixedness. 
The wave feature vanishes at \(k=1/2\) and revives as \(k\) further 
increases, whereas the mixedness shows the opposite trend. 
The particle feature remains independent of the damping coefficient. 
Thus, the calculated quantities are consistent with the 
wave-particle-mixedness relation.

\subsection{C. Bit flip channel}

The bit flip channel differs from the PD and PF channels because it 
exchanges the computational basis states and redistributes the diagonal 
populations. 
For the reduced states considered here, the wave feature has no explicit 
dependence on the damping coefficient \(k\), whereas the particle feature 
and mixedness are directly modified by the \(k\)-dependent population 
redistribution. 
The noisy reduced states generated by the BF channel, together with the 
corresponding wave feature, particle feature, and mixedness, are given in 
Appendix~B.

\begin{figure*}[!t]
\centering

\makebox[\textwidth][c]{%
\includegraphics[width=1.04\textwidth]{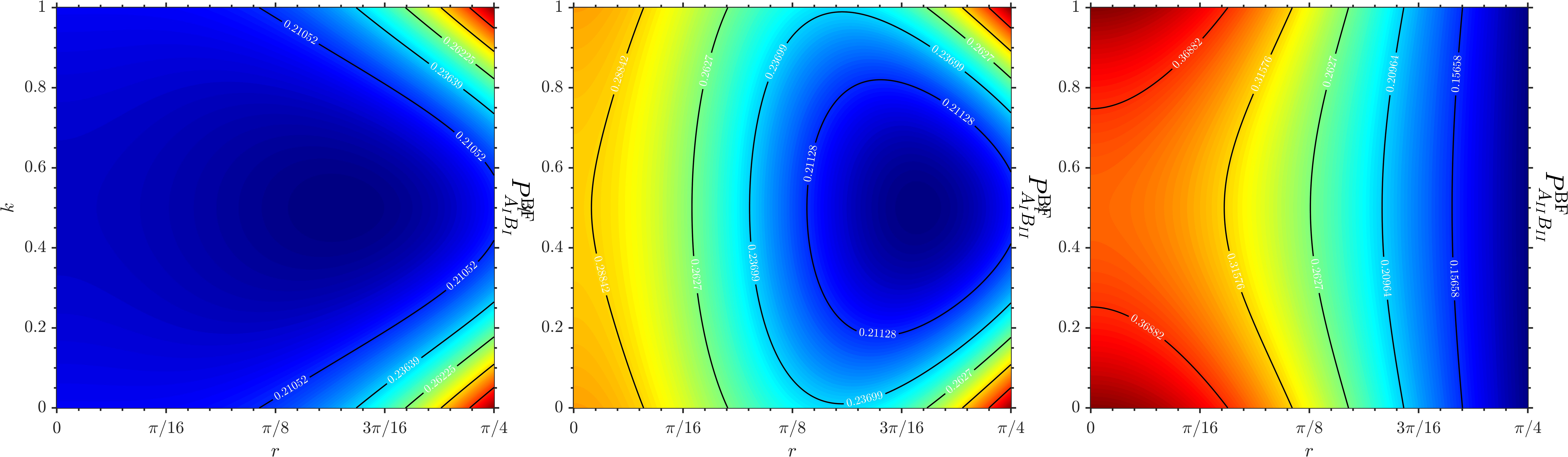}
}

\vspace{1mm}
\centerline{\textbf{(a)}}

\vspace{3mm}

\makebox[\textwidth][c]{%
\includegraphics[width=1.04\textwidth]{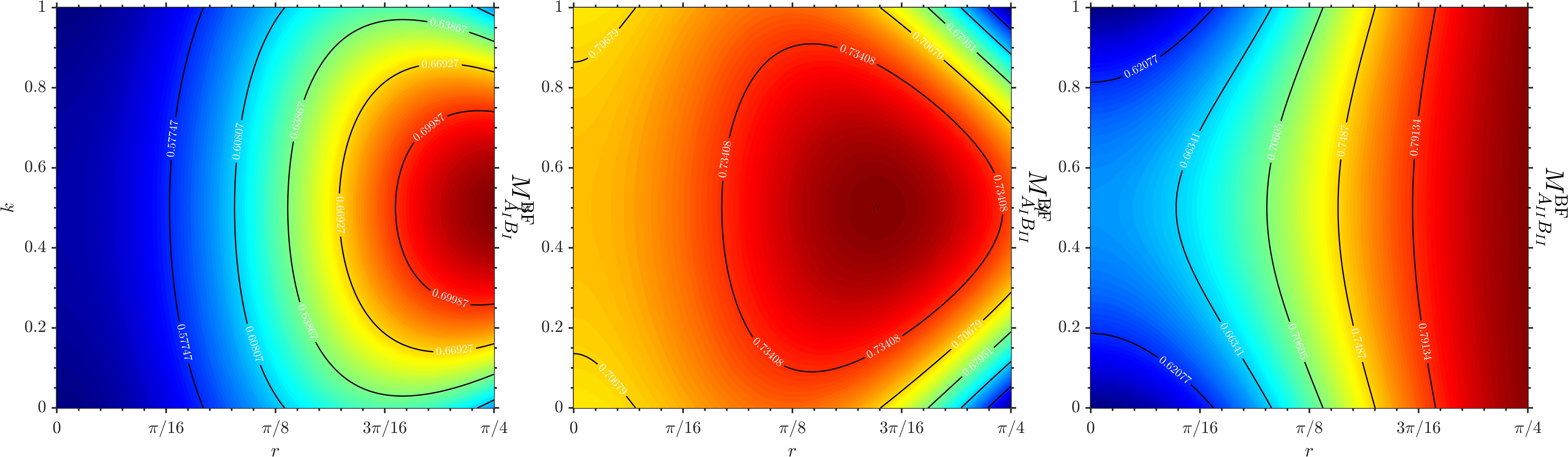}
}

\vspace{1mm}
\centerline{\textbf{(b)}}

\vspace{2mm}

\caption{
Particle feature and mixedness under the bit flip channel 
as functions of the Hawking parameter \(r\) 
and the damping coefficient \(k\). 
Panel (a) shows the particle feature 
\(\mathcal{P}^{\mathrm{BF}}\), 
while panel (b) shows the mixedness 
\(\mathcal{M}^{\mathrm{BF}}\). 
In each panel, the three subplots correspond to 
\(\rho^{\mathrm{BF}}_{A_I B_I}\), 
\(\rho^{\mathrm{BF}}_{A_I B_{II}}\), 
and 
\(\rho^{\mathrm{BF}}_{A_{II}B_{II}}\). 
The wave feature is not plotted because it has no explicit 
dependence on \(k\). 
For \(r_a=r_b=r\), 
\(\rho^{\mathrm{BF}}_{A_{II}B_I}\) 
has the same behavior as 
\(\rho^{\mathrm{BF}}_{A_I B_{II}}\).
}
\label{fig:BF_all}
\end{figure*}

Equations~(\ref{eq:BF_W_all_C})--(\ref{eq:BF_M_all_C}) 
show that the BF channel mainly reshapes the triality distribution 
through the particle feature and mixedness. 
Since the wave feature contains no explicit \(k\)-dependent factor, 
it is not plotted in Fig.~\ref{fig:BF_all}. 
For \(r_a=r_b=r\), 
the reduced states 
\(\rho^{\mathrm{BF}}_{A_I B_{II}}\) 
and 
\(\rho^{\mathrm{BF}}_{A_{II}B_I}\) 
have the same behavior, 
so only one of them is shown.

Fig.~\ref{fig:BF_all}(a) shows the particle feature 
of the reduced states after the BF channel 
as functions of the Hawking parameter \(r\) 
and the damping coefficient \(k\). 
It can be seen that 
\(\rho^{\mathrm{BF}}_{A_I B_I}\), 
\(\rho^{\mathrm{BF}}_{A_I B_{II}}\), 
and 
\(\rho^{\mathrm{BF}}_{A_{II}B_{II}}\) 
all exhibit a clear two-parameter dependence. 
This behavior is in sharp contrast to the PD and PF channels, 
where the particle feature is independent of 
the noise strength. 
The difference originates from the bit-flip operation, 
which exchanges the computational basis states 
and redistributes the diagonal populations. 
From Eq.~(\ref{eq:BF_P_all_C}), 
one finds that the particle feature contains 
a \(k(k-1)\)-dependent contribution. 
Since \(k(k-1)\) reaches its minimum at \(k=1/2\), 
the particle feature is generally suppressed 
in the intermediate-noise region. 
Meanwhile, the Hawking parameter changes the population weights 
associated with the physically accessible and inaccessible sectors, 
leading to different contour structures for the three reduced states.

Fig.~\ref{fig:BF_all}(b) presents the corresponding mixedness. 
The contour profiles show a behavior complementary to that of 
the particle feature in Fig.~\ref{fig:BF_all}(a). 
In the parameter regions where the particle feature is weakened, 
the mixedness is enhanced, 
which indicates that the bit-flip-induced population redistribution 
increases the statistical uncertainty of the reduced state. 
This complementary behavior is especially evident near 
intermediate values of \(k\), 
where the effect of the \(k(k-1)\) term is most pronounced. 
By contrast, the wave feature is not plotted because 
Eq.~(\ref{eq:BF_W_all_C}) contains no explicit dependence on \(k\). 
Therefore, the BF channel mainly redistributes the triality weight 
between the particle feature and mixedness, 
rather than directly suppressing the wave feature.

Substituting 
Eqs.~(\ref{eq:BF_W_all_C}), 
(\ref{eq:BF_P_all_C}), 
and 
(\ref{eq:BF_M_all_C}) 
into Eq.~(\ref{eq:triality}) also gives 
\(\mathcal{P}(\rho)+\mathcal{W}(\rho)+\mathcal{M}(\rho)=1\). 
Therefore, the BF channel modifies the wave-particle-mixedness 
distribution mainly through the particle feature and mixedness. 
Unlike the PD and PF channels, its dominant effect originates from the 
redistribution of diagonal populations. 
The wave feature remains independent of the damping coefficient, whereas 
the particle feature and mixedness show clear two-parameter dependence on 
\(r\) and \(k\). 
Thus, the calculated quantities are consistent with 
Eq.~(\ref{eq:triality}) after substitution.

\subsection{D. Comparison of the three noisy channels}

The results obtained above show that the three noisy channels modify 
the wave feature, particle feature, and mixedness in qualitatively 
different ways. 
This difference originates from their distinct actions on the diagonal 
and off-diagonal elements of the reduced density matrix.

For the phase damping channel, the off-diagonal elements are suppressed 
by the factor \((1-k)\), so the wave feature is monotonically reduced 
by the factor \((1-k)^2\). 
Since the diagonal populations remain unchanged, the particle feature 
has no explicit dependence on the damping coefficient. 
The suppression of the wave feature is accompanied by an increase in 
mixedness, leading to a monotonic redistribution between the wave 
feature and mixedness.

For the phase flip channel, the off-diagonal elements acquire the 
factor \((1-2k)\). 
Accordingly, the wave feature is governed by \((1-2k)^2\), vanishes at 
\(k=1/2\), and increases again as \(k\) approaches unity. 
The particle feature remains independent of the damping coefficient, 
whereas the mixedness exhibits the opposite symmetric behavior. 
Thus, the phase flip channel produces a symmetric redistribution between 
the wave feature and mixedness.

The bit flip channel behaves differently from the phase damping and 
phase flip channels. 
It exchanges the computational basis states and directly redistributes 
the diagonal populations. 
Consequently, the particle feature and mixedness acquire explicit 
dependence on the damping coefficient, whereas the wave feature is not 
directly suppressed by \(k\). 
This indicates that the bit flip channel mainly reshapes the 
wave-particle-mixedness distribution through diagonal population 
redistribution.

These comparisons show that environmental noise does not modify the 
three quantities in a universal manner. 
Instead, each channel produces a characteristic redistribution pattern 
determined by its action on the diagonal or off-diagonal elements of 
the reduced density matrix. 
After \(\mathcal{W}\), \(\mathcal{P}\), and \(\mathcal{M}\) are calculated 
from the noisy reduced states, their substitution into 
Eq.~(\ref{eq:triality}) provides a consistency check for the obtained 
redistribution results.

\section{V. Conclusions}

We have investigated the redistribution of the wave feature, particle feature, and mixedness for a two-qubit isotropic Dirac state in Schwarzschild spacetime. 
After the Dirac field modes are separated into the physically accessible exterior region and the physically inaccessible interior region, four noninertial reduced states are obtained. 
For each reduced state, the three quantities are calculated explicitly from the corresponding density matrix. 
The results show that Hawking radiation changes their relative weights among different horizon sectors in a state dependent manner.

In the absence of environmental noise, the redistribution pattern depends strongly on the horizon region under consideration. 
For the physically accessible reduced state, the wave feature decreases as the Hawking parameter increases. 
For reduced states involving Region-II modes, the inaccessible-sector contribution becomes more important and leads to different behaviors of the wave feature, particle feature, and mixedness. 
Thus, Hawking radiation does not simply degrade these quantities; instead, it reorganizes them between the exterior and interior region.

The influence of environmental decoherence has also been analyzed through the phase-damping, phase-flip, and bit-flip channels. 
The phase-damping channel monotonically suppresses the wave feature and enhances mixedness, while leaving the particle feature independent of the damping coefficient. 
The phase-flip channel produces a symmetric death-and-revival behavior of the wave feature and an opposite variation of mixedness. 
The bit-flip channel mainly modifies the particle feature and mixedness through the redistribution of diagonal populations. 
These results show that different noise mechanisms reshape the wave-particle-mixedness distribution in qualitatively different ways.

The triality relation has been used as a consistency check for the quantities obtained from the reduced density matrices. 
Direct substitution shows that the horizon and channel dependent redistributions occur within the expected total wave-particle-mixedness balance. 
This provides a useful perspective for understanding how quantum complementarity is modified in curved spacetime and noisy relativistic quantum systems.

Several extensions remain open. 
The present work is restricted to a static Schwarzschild black hole, 
fermionic Dirac modes, two-qubit isotropic states, 
and three representative noisy channels. 
It would be interesting to extend the analysis to rotating black holes, 
bosonic fields, multipartite states, and more general open-system 
dynamics. 
Such extensions may further clarify how wave-particle-mixedness 
triality behaves in more realistic relativistic quantum-information 
scenarios.

\section*{Acknowledgments}

This work is supported by the Natural Science Foundation of Hainan Province under Grant No. 125RC744; the China Scholarship Council (CSC).

\appendix

\allowdisplaybreaks[4]
\setlength{\abovedisplayskip}{3pt}
\setlength{\belowdisplayskip}{3pt}
\setlength{\abovedisplayshortskip}{2pt}
\setlength{\belowdisplayshortskip}{2pt}

\setcounter{equation}{0}
\renewcommand{\theequation}{A\arabic{equation}}

\section{Appendix A: Two-body density matrices for isotropic states in Schwarzschild spacetime}

In the computational basis $\{\ket{00},\ket{01},\ket{10},\ket{11}\}$, it can be written as
\begin{equation}
\label{eq:rho_AIBI_matrix}
\rho_{A_I B_I}
=
\left(
\begin{array}{cccc}
\rho_{11}^{(1)} & 0 & 0 & 0 \\
0 & \rho_{22}^{(1)} & \rho_{23}^{(1)} & 0 \\
0 & \rho_{32}^{(1)} & \rho_{33}^{(1)} & 0 \\
0 & 0 & 0 & \rho_{44}^{(1)}
\end{array}
\right),
\end{equation}
where the nonzero matrix elements are
\begin{align}
\rho_{11}^{(1)}
&=
\frac{1}{4}
\Bigl(
1-\sin^2 r_a-\sin^2 r_b
+\sin^2 r_a\sin^2 r_b
\notag\\
&\quad
-\alpha\cos^2 r_a\cos^2 r_b
\Bigr),
\notag\\
\rho_{22}^{(1)}
&=
\frac{1}{4}
\Bigl(
1-\sin^2 r_a+\sin^2 r_b
-\sin^2 r_a\sin^2 r_b
\notag\\
&\quad
+\alpha\cos^2 r_a\cos^2 r_b
\Bigr),
\notag\\
\rho_{33}^{(1)}
&=
\frac{1}{4}
\Bigl(
1+\sin^2 r_a-\sin^2 r_b
-\sin^2 r_a\sin^2 r_b
\notag\\
&\quad
+\alpha\cos^2 r_a\cos^2 r_b
\Bigr),
\notag\\
\rho_{44}^{(1)}
&=
\frac{1}{4}
\Bigl(
1+\sin^2 r_a+\sin^2 r_b
+\sin^2 r_a\sin^2 r_b
\notag\\
&\quad
-\alpha\cos^2 r_a\cos^2 r_b
\Bigr),
\notag\\
\rho_{23}^{(1)}
&=
\rho_{32}^{(1)}
=
\frac{\alpha}{2}\cos r_a\cos r_b .
\label{eq:rho_AIBI_elements}
\end{align}

For the mixed-region reduced state \(\rho_{A_I B_{II}}\), in which Alice's mode remains outside the event horizon while Bob's mode lies inside the event horizon, the reduced density matrix is
\begin{equation}
\rho_{A_I B_{II}} =
\begin{pmatrix}
\rho_{11}^{(2)} & 0 & 0 & \rho_{14}^{(2)} \\
0 & \rho_{22}^{(2)} & 0 & 0 \\
0 & 0 & \rho_{33}^{(2)} & 0 \\
\rho_{41}^{(2)} & 0 & 0 & \rho_{44}^{(2)}
\end{pmatrix},
\end{equation}
where
\begin{align}
\rho_{11}^{(2)}
&=
\frac14
\Bigl(
1-\sin^2 r_a+\cos^2 r_b
\notag\\
&\quad
-\sin^2 r_a\cos^2 r_b
+\alpha\cos^2 r_a\sin^2 r_b
\Bigr),
\\
\rho_{22}^{(2)}
&=
\frac14
\Bigl(
1-\sin^2 r_a-\cos^2 r_b
\notag\\
&\quad
+\sin^2 r_a\cos^2 r_b
-\alpha\cos^2 r_a\sin^2 r_b
\Bigr),
\\
\rho_{33}^{(2)}
&=
\frac14
\Bigl(
1+\sin^2 r_a+\cos^2 r_b
\notag\\
&\quad
-\sin^2 r_a\cos^2 r_b
-\alpha\cos^2 r_a\sin^2 r_b
\Bigr),
\\
\rho_{44}^{(2)}
&=
\frac14
\Bigl(
1+\sin^2 r_a-\cos^2 r_b
\notag\\
&\quad
+\sin^2 r_a\cos^2 r_b
+\alpha\cos^2 r_a\sin^2 r_b
\Bigr),
\\
\rho_{14}^{(2)}
&=
\rho_{41}^{(2)}
=
\frac{\alpha}{2}\cos r_a\sin r_b .
\end{align}

For the mixed-region reduced state \(\rho_{A_{II} B_I}\), in which Alice's mode lies inside the event horizon while Bob's mode remains outside the event horizon, one obtains
\begin{equation}
\rho_{A_{II} B_I} =
\begin{pmatrix}
\rho_{11}^{(3)} & 0 & 0 & \rho_{14}^{(3)} \\
0 & \rho_{22}^{(3)} & 0 & 0 \\
0 & 0 & \rho_{33}^{(3)} & 0 \\
\rho_{41}^{(3)} & 0 & 0 & \rho_{44}^{(3)}
\end{pmatrix},
\end{equation}
with
\begin{align}
\rho_{11}^{(3)}
&=
\frac14
\Bigl(
1+\cos^2 r_a-\sin^2 r_b
\notag\\
&\quad
+\alpha\sin^2 r_a\cos^2 r_b
-\cos^2 r_a\sin^2 r_b
\Bigr),
\\
\rho_{22}^{(3)}
&=
\frac14
\Bigl(
1+\cos^2 r_a+\sin^2 r_b
\notag\\
&\quad
-\alpha\sin^2 r_a\cos^2 r_b
+\cos^2 r_a\sin^2 r_b
\Bigr),
\\
\rho_{33}^{(3)}
&=
\frac14
\Bigl(
1-\cos^2 r_a-\sin^2 r_b
\notag\\
&\quad
-\alpha\sin^2 r_a\cos^2 r_b
+\cos^2 r_a\sin^2 r_b
\Bigr),
\\
\rho_{44}^{(3)}
&=
\frac14
\Bigl(
1-\cos^2 r_a+\sin^2 r_b
\notag\\
&\quad
+\alpha\sin^2 r_a\cos^2 r_b
-\cos^2 r_a\sin^2 r_b
\Bigr),
\\
\rho_{14}^{(3)}
&=
\rho_{41}^{(3)}
=
\frac{\alpha}{2}\sin r_a\cos r_b .
\end{align}

When both modes are located in the physically inaccessible region inside the event horizon, the corresponding density matrix is
\begin{equation}
\rho_{A_{II} B_{II}} =
\begin{pmatrix}
\rho_{11}^{(4)} & 0 & 0 & 0 \\
0 & \rho_{22}^{(4)} & \rho_{23}^{(4)} & 0 \\
0 & \rho_{32}^{(4)} & \rho_{33}^{(4)} & 0 \\
0 & 0 & 0 & \rho_{44}^{(4)}
\end{pmatrix},
\end{equation}
where
\begin{align}
\rho_{11}^{(4)}
&=
\frac14
\Bigl(
1+\cos^2 r_a+\cos^2 r_b
\notag\\
&\quad
+\cos^2 r_a\cos^2 r_b
-\alpha\sin^2 r_a\sin^2 r_b
\Bigr),
\\
\rho_{22}^{(4)}
&=
\frac14
\Bigl(
1+\cos^2 r_a-\cos^2 r_b
\notag\\
&\quad
-\cos^2 r_a\cos^2 r_b
+\alpha\sin^2 r_a\sin^2 r_b
\Bigr),
\\
\rho_{33}^{(4)}
&=
\frac14
\Bigl(
1-\cos^2 r_a+\cos^2 r_b
\notag\\
&\quad
-\cos^2 r_a\cos^2 r_b
+\alpha\sin^2 r_a\sin^2 r_b
\Bigr),
\\
\rho_{44}^{(4)}
&=
\frac14
\Bigl(
1-\cos^2 r_a-\cos^2 r_b
\notag\\
&\quad
+\cos^2 r_a\cos^2 r_b
-\alpha\sin^2 r_a\sin^2 r_b
\Bigr),
\\
\rho_{23}^{(4)}
&=
\rho_{32}^{(4)}
=
\frac{\alpha}{2}\sin r_a\sin r_b .
\end{align}

These matrix elements are used to calculate the wave feature, 
particle feature, and mixedness in the main text.

\vspace{-0.5\baselineskip}

\setcounter{equation}{0}
\renewcommand{\theequation}{B\arabic{equation}}

These matrix elements are used to calculate the wave feature, 
particle feature, and mixedness in the main text.

\begin{widetext}

\setcounter{equation}{0}
\renewcommand{\theequation}{B\arabic{equation}}

\section{Appendix B: Noisy reduced states and triality quantities}
\label{app:noise_channel_results}

This appendix collects the noisy reduced states and the corresponding 
wave feature, particle feature, and mixedness for the phase damping, 
phase flip, and bit flip channels. 
The results are organized according to the three noisy channels.

For compactness, a generic noiseless reduced state is written as
\begin{equation}
\rho_X
=
\begin{pmatrix}
D_1^X & 0 & 0 & 0 \\
0 & D_2^X & E_X & 0 \\
0 & E_X & D_3^X & 0 \\
0 & 0 & 0 & D_4^X
\end{pmatrix},
\label{eq:generic_X_state}
\end{equation}
where
\[
X
\in
\{
A_I B_I,
A_I B_{II},
A_{II}B_I,
A_{II}B_{II}
\}.
\]
Here \(D_i^X\) are the diagonal entries and \(E_X\) is 
the nonzero off-diagonal element of the corresponding reduced state 
given in Appendix~A. 
For all noisy states, the mixedness is calculated from
\[
\mathcal{M}(\rho_X^\mu)
=
1
-
\mathrm{tr}
\left[
(\rho_X^\mu)^2
\right],
\qquad
\mu
=
\mathrm{PD},
\mathrm{PF},
\mathrm{BF}.
\]

\subsection{A. Phase damping channel}
\label{app:PD_channel}

For the phase damping channel, the diagonal entries are unchanged, 
whereas the nonzero off-diagonal element is multiplied by \(1-k\). 
Thus, the PD-channel reduced state is
\begin{equation}
\rho_X^{\mathrm{PD}}
=
\begin{pmatrix}
D_1^X & 0 & 0 & 0 \\
0 & D_2^X & (1-k)E_X & 0 \\
0 & (1-k)E_X & D_3^X & 0 \\
0 & 0 & 0 & D_4^X
\end{pmatrix}.
\label{eq:PD_noisy_matrix_general}
\end{equation}

For the four reduced states, the wave features under the PD channel are
\begin{equation}
\begin{alignedat}{2}
\mathcal{W}^{\mathrm{PD}}_{A_I B_I}
&=
(1-k)^2
\frac{\alpha^2}{2}
\cos^2 r_a
\cos^2 r_b ,
\qquad
&
\mathcal{W}^{\mathrm{PD}}_{A_I B_{II}}
&=
(1-k)^2
\frac{\alpha^2}{2}
\cos^2 r_a
\sin^2 r_b ,
\\[7pt]
\mathcal{W}^{\mathrm{PD}}_{A_{II}B_I}
&=
(1-k)^2
\frac{\alpha^2}{2}
\sin^2 r_a
\cos^2 r_b ,
\qquad
&
\mathcal{W}^{\mathrm{PD}}_{A_{II}B_{II}}
&=
(1-k)^2
\frac{\alpha^2}{2}
\sin^2 r_a
\sin^2 r_b .
\end{alignedat}
\label{eq:PD_W_all}
\end{equation}

The corresponding particle features are
\begin{equation}
\begin{aligned}
\mathcal{P}^{\mathrm{PD}}_{A_I B_I}
&=
\frac{1}{4}
\Bigl[
1+\sin^4 r_a+\sin^4 r_b
+
\bigl(
\sin^2 r_a\sin^2 r_b
-
\alpha\cos^2 r_a\cos^2 r_b
\bigr)^2
\Bigr],
\\[8pt]
\mathcal{P}^{\mathrm{PD}}_{A_I B_{II}}
&=
\frac{1}{4}
\Bigl[
1+\sin^4 r_a+\cos^4 r_b
+
\bigl(
\alpha\cos^2 r_a\sin^2 r_b
-
\sin^2 r_a\cos^2 r_b
\bigr)^2
\Bigr],
\\[8pt]
\mathcal{P}^{\mathrm{PD}}_{A_{II}B_I}
&=
\frac{1}{4}
\Bigl[
1+\cos^4 r_a+\sin^4 r_b
+
\bigl(
\alpha\sin^2 r_a\cos^2 r_b
-
\cos^2 r_a\sin^2 r_b
\bigr)^2
\Bigr],
\\[8pt]
\mathcal{P}^{\mathrm{PD}}_{A_{II}B_{II}}
&=
\frac{1}{4}
\Bigl[
1+\cos^4 r_a+\cos^4 r_b
+
\bigl(
\cos^2 r_a\cos^2 r_b
-
\alpha\sin^2 r_a\sin^2 r_b
\bigr)^2
\Bigr].
\end{aligned}
\label{eq:PD_P_all}
\end{equation}

The mixednesses obtained from the linear entropy are
\begin{equation}
\begin{aligned}
\mathcal{M}^{\mathrm{PD}}_{A_I B_I}
&=
\frac{3}{4}
-
\frac{1}{4}
\Bigl[
\sin^4 r_a+\sin^4 r_b
+
\bigl(
\sin^2 r_a\sin^2 r_b
-
\alpha\cos^2 r_a\cos^2 r_b
\bigr)^2
\Bigr]
-
(1-k)^2
\frac{\alpha^2}{2}
\cos^2 r_a\cos^2 r_b,
\\[8pt]
\mathcal{M}^{\mathrm{PD}}_{A_I B_{II}}
&=
\frac{3}{4}
-
\frac{1}{4}
\Bigl[
\sin^4 r_a+\cos^4 r_b
+
\bigl(
\alpha\cos^2 r_a\sin^2 r_b
-
\sin^2 r_a\cos^2 r_b
\bigr)^2
\Bigr]
-
(1-k)^2
\frac{\alpha^2}{2}
\cos^2 r_a\sin^2 r_b,
\\[8pt]
\mathcal{M}^{\mathrm{PD}}_{A_{II}B_I}
&=
\frac{3}{4}
-
\frac{1}{4}
\Bigl[
\cos^4 r_a+\sin^4 r_b
+
\bigl(
\alpha\sin^2 r_a\cos^2 r_b
-
\cos^2 r_a\sin^2 r_b
\bigr)^2
\Bigr]
-
(1-k)^2
\frac{\alpha^2}{2}
\sin^2 r_a\cos^2 r_b,
\\[8pt]
\mathcal{M}^{\mathrm{PD}}_{A_{II}B_{II}}
&=
\frac{3}{4}
-
\frac{1}{4}
\Bigl[
\cos^4 r_a+\cos^4 r_b
+
\bigl(
\cos^2 r_a\cos^2 r_b
-
\alpha\sin^2 r_a\sin^2 r_b
\bigr)^2
\Bigr]
-
(1-k)^2
\frac{\alpha^2}{2}
\sin^2 r_a\sin^2 r_b.
\end{aligned}
\label{eq:PD_M_all}
\end{equation}

\subsection{B. Phase flip channel}
\label{app:PF_channel}

For the phase flip channel, the diagonal entries are unchanged, 
whereas the nonzero off-diagonal element is multiplied by \(1-2k\). 
Thus, the PF-channel reduced state is
\begin{equation}
\rho_X^{\mathrm{PF}}
=
\begin{pmatrix}
D_1^X & 0 & 0 & 0 \\
0 & D_2^X & (1-2k)E_X & 0 \\
0 & (1-2k)E_X & D_3^X & 0 \\
0 & 0 & 0 & D_4^X
\end{pmatrix}.
\label{eq:PF_noisy_matrix_general}
\end{equation}

For the four reduced states, the wave features under the PF channel are
\begin{equation}
\begin{alignedat}{2}
\mathcal{W}^{\mathrm{PF}}_{A_I B_I}
&=
(1-2k)^2
\frac{\alpha^2}{2}
\cos^2 r_a
\cos^2 r_b ,
\qquad
&
\mathcal{W}^{\mathrm{PF}}_{A_I B_{II}}
&=
(1-2k)^2
\frac{\alpha^2}{2}
\cos^2 r_a
\sin^2 r_b ,
\\[7pt]
\mathcal{W}^{\mathrm{PF}}_{A_{II}B_I}
&=
(1-2k)^2
\frac{\alpha^2}{2}
\sin^2 r_a
\cos^2 r_b ,
\qquad
&
\mathcal{W}^{\mathrm{PF}}_{A_{II}B_{II}}
&=
(1-2k)^2
\frac{\alpha^2}{2}
\sin^2 r_a
\sin^2 r_b .
\end{alignedat}
\label{eq:PF_W_all_C}
\end{equation}

The corresponding particle features are
\begin{equation}
\begin{aligned}
\mathcal{P}^{\mathrm{PF}}_{A_I B_I}
&=
\frac{1}{4}
\Bigl[
1
+
\sin^4 r_a
+
\sin^4 r_b
+
\bigl(
\sin^2 r_a
\sin^2 r_b
-
\alpha
\cos^2 r_a
\cos^2 r_b
\bigr)^2
\Bigr],
\\[8pt]
\mathcal{P}^{\mathrm{PF}}_{A_I B_{II}}
&=
\frac{1}{4}
\Bigl[
1
+
\sin^4 r_a
+
\cos^4 r_b
+
\bigl(
\alpha
\cos^2 r_a
\sin^2 r_b
-
\sin^2 r_a
\cos^2 r_b
\bigr)^2
\Bigr],
\\[8pt]
\mathcal{P}^{\mathrm{PF}}_{A_{II}B_I}
&=
\frac{1}{4}
\Bigl[
1
+
\cos^4 r_a
+
\sin^4 r_b
+
\bigl(
\alpha
\sin^2 r_a
\cos^2 r_b
-
\cos^2 r_a
\sin^2 r_b
\bigr)^2
\Bigr],
\\[8pt]
\mathcal{P}^{\mathrm{PF}}_{A_{II}B_{II}}
&=
\frac{1}{4}
\Bigl[
1
+
\cos^4 r_a
+
\cos^4 r_b
+
\bigl(
\cos^2 r_a
\cos^2 r_b
-
\alpha
\sin^2 r_a
\sin^2 r_b
\bigr)^2
\Bigr].
\end{aligned}
\label{eq:PF_P_all_C}
\end{equation}

The mixednesses obtained from the linear entropy are
\begin{equation}
\begin{aligned}
\mathcal{M}^{\mathrm{PF}}_{A_I B_I}
&=
\frac{3}{4}
-
\frac{1}{4}
\Bigl[
\sin^4 r_a+\sin^4 r_b
+
\bigl(
\sin^2 r_a\sin^2 r_b
-
\alpha\cos^2 r_a\cos^2 r_b
\bigr)^2
\Bigr]
-
(1-2k)^2
\frac{\alpha^2}{2}
\cos^2 r_a\cos^2 r_b,
\\[8pt]
\mathcal{M}^{\mathrm{PF}}_{A_I B_{II}}
&=
\frac{3}{4}
-
\frac{1}{4}
\Bigl[
\sin^4 r_a+\cos^4 r_b
+
\bigl(
\alpha\cos^2 r_a\sin^2 r_b
-
\sin^2 r_a\cos^2 r_b
\bigr)^2
\Bigr]
-
(1-2k)^2
\frac{\alpha^2}{2}
\cos^2 r_a\sin^2 r_b,
\\[8pt]
\mathcal{M}^{\mathrm{PF}}_{A_{II}B_I}
&=
\frac{3}{4}
-
\frac{1}{4}
\Bigl[
\cos^4 r_a+\sin^4 r_b
+
\bigl(
\alpha\sin^2 r_a\cos^2 r_b
-
\cos^2 r_a\sin^2 r_b
\bigr)^2
\Bigr]
-
(1-2k)^2
\frac{\alpha^2}{2}
\sin^2 r_a\cos^2 r_b,
\\[8pt]
\mathcal{M}^{\mathrm{PF}}_{A_{II}B_{II}}
&=
\frac{3}{4}
-
\frac{1}{4}
\Bigl[
\cos^4 r_a+\cos^4 r_b
+
\bigl(
\cos^2 r_a\cos^2 r_b
-
\alpha\sin^2 r_a\sin^2 r_b
\bigr)^2
\Bigr]
-
(1-2k)^2
\frac{\alpha^2}{2}
\sin^2 r_a\sin^2 r_b.
\end{aligned}
\label{eq:PF_M_all_C}
\end{equation}

\subsection{C. Bit flip channel}
\label{app:BF_channel}

For the bit flip channel, the off-diagonal contribution to the 
wave feature has no explicit \(k\)-dependent damping factor. 
The channel instead redistributes the diagonal populations. 
The corresponding noisy state can still be written in an X-shaped form,
\begin{equation}
\rho_X^{\mathrm{BF}}
=
\begin{pmatrix}
\widetilde{D}_1^X(k) & 0 & 0 & 0 \\
0 & \widetilde{D}_2^X(k) & E_X & 0 \\
0 & E_X & \widetilde{D}_3^X(k) & 0 \\
0 & 0 & 0 & \widetilde{D}_4^X(k)
\end{pmatrix},
\label{eq:BF_noisy_matrix_general}
\end{equation}
where \(\widetilde{D}_i^X(k)\) denotes the diagonal population 
after the BF-channel redistribution.

For the four reduced states, the wave features under the BF channel are
\begin{equation}
\begin{alignedat}{2}
\mathcal{W}^{\mathrm{BF}}_{A_I B_I}
&=
\frac{\alpha^2}{2}
\cos^2 r_a
\cos^2 r_b ,
\qquad
&
\mathcal{W}^{\mathrm{BF}}_{A_I B_{II}}
&=
\frac{\alpha^2}{2}
\cos^2 r_a
\sin^2 r_b ,
\\[7pt]
\mathcal{W}^{\mathrm{BF}}_{A_{II}B_I}
&=
\frac{\alpha^2}{2}
\sin^2 r_a
\cos^2 r_b ,
\qquad
&
\mathcal{W}^{\mathrm{BF}}_{A_{II}B_{II}}
&=
\frac{\alpha^2}{2}
\sin^2 r_a
\sin^2 r_b .
\end{alignedat}
\label{eq:BF_W_all_C}
\end{equation}

The corresponding particle features are
\begin{equation}
\begin{aligned}
\mathcal{P}^{\mathrm{BF}}_{A_I B_I}
&=
\frac{1}{4}
\Bigl[
1+\sin^4 r_a+\sin^4 r_b
+
\bigl(
\sin^2 r_a\sin^2 r_b
-
\alpha\cos^2 r_a\cos^2 r_b
\bigr)^2
+
4k(k-1)
\bigl(
\sin^4 r_a+\sin^4 r_b
\bigr)
\Bigr],
\\[8pt]
\mathcal{P}^{\mathrm{BF}}_{A_I B_{II}}
&=
\frac{1}{4}
\Bigl[
1+\sin^4 r_a+\cos^4 r_b
+
\bigl(
\alpha\cos^2 r_a\sin^2 r_b
-
\sin^2 r_a\cos^2 r_b
\bigr)^2
+
4k(k-1)
\bigl(
\sin^4 r_a+\cos^4 r_b
\bigr)
\Bigr],
\\[8pt]
\mathcal{P}^{\mathrm{BF}}_{A_{II}B_I}
&=
\frac{1}{4}
\Bigl[
1+\cos^4 r_a+\sin^4 r_b
+
\bigl(
\alpha\sin^2 r_a\cos^2 r_b
-
\cos^2 r_a\sin^2 r_b
\bigr)^2
+
4k(k-1)
\bigl(
\cos^4 r_a+\sin^4 r_b
\bigr)
\Bigr],
\\[8pt]
\mathcal{P}^{\mathrm{BF}}_{A_{II}B_{II}}
&=
\frac{1}{4}
\Bigl[
1+\cos^4 r_a+\cos^4 r_b
+
\bigl(
\cos^2 r_a\cos^2 r_b
-
\alpha\sin^2 r_a\sin^2 r_b
\bigr)^2
+
4k(k-1)
\bigl(
\cos^4 r_a+\cos^4 r_b
\bigr)
\Bigr].
\end{aligned}
\label{eq:BF_P_all_C}
\end{equation}

The mixednesses obtained from the linear entropy are
\begin{equation}
\begin{aligned}
\mathcal{M}^{\mathrm{BF}}_{A_I B_I}
&=
\frac{3}{4}
-
\frac{1}{4}
\Bigl[
\sin^4 r_a+\sin^4 r_b
+
\bigl(
\sin^2 r_a\sin^2 r_b
-
\alpha\cos^2 r_a\cos^2 r_b
\bigr)^2
+
4k(k-1)
\bigl(
\sin^4 r_a+\sin^4 r_b
\bigr)
\Bigr]
\\
&\quad
-
\frac{\alpha^2}{2}
\cos^2 r_a\cos^2 r_b,
\\[8pt]
\mathcal{M}^{\mathrm{BF}}_{A_I B_{II}}
&=
\frac{3}{4}
-
\frac{1}{4}
\Bigl[
\sin^4 r_a+\cos^4 r_b
+
\bigl(
\alpha\cos^2 r_a\sin^2 r_b
-
\sin^2 r_a\cos^2 r_b
\bigr)^2
+
4k(k-1)
\bigl(
\sin^4 r_a+\cos^4 r_b
\bigr)
\Bigr]
\\
&\quad
-
\frac{\alpha^2}{2}
\cos^2 r_a\sin^2 r_b,
\\[8pt]
\mathcal{M}^{\mathrm{BF}}_{A_{II}B_I}
&=
\frac{3}{4}
-
\frac{1}{4}
\Bigl[
\cos^4 r_a+\sin^4 r_b
+
\bigl(
\alpha\sin^2 r_a\cos^2 r_b
-
\cos^2 r_a\sin^2 r_b
\bigr)^2
+
4k(k-1)
\bigl(
\cos^4 r_a+\sin^4 r_b
\bigr)
\Bigr]
\\
&\quad
-
\frac{\alpha^2}{2}
\sin^2 r_a\cos^2 r_b,
\\[8pt]
\mathcal{M}^{\mathrm{BF}}_{A_{II}B_{II}}
&=
\frac{3}{4}
-
\frac{1}{4}
\Bigl[
\cos^4 r_a+\cos^4 r_b
+
\bigl(
\cos^2 r_a\cos^2 r_b
-
\alpha\sin^2 r_a\sin^2 r_b
\bigr)^2
+
4k(k-1)
\bigl(
\cos^4 r_a+\cos^4 r_b
\bigr)
\Bigr]
\\
&\quad
-
\frac{\alpha^2}{2}
\sin^2 r_a\sin^2 r_b .
\end{aligned}
\label{eq:BF_M_all_C}
\end{equation}

The above expressions show that the BF channel affects the particle 
feature and mixedness through the \(k(k-1)\)-dependent diagonal 
population redistribution, whereas the wave feature has no explicit 
dependence on \(k\).

\end{widetext}

\end{document}